\documentclass[aps,amsfonts,pra,preprint,singlecolumn,showpacs]{revtex4}
\usepackage{epsfig,amsmath,amssymb,bm,epsf,graphicx}

\def\ket#1{\vert#1\rangle}
\def\ketbra#1{\vert#1\rangle\langle#1\vert}

\def\Longarrow{\protect\@lra}
\def\@lra{\relbar\joinrel\relbar\joinrel\relbar\joinrel%
          \relbar\joinrel\rightarrow}

\def\Chi{X}
\begin{document}
\title{Creating  multi-photon polarization bound-entangled states}

\author{Tzu-Chieh Wei}
\affiliation{Department of Physics and Astronomy, University of
British Columbia, Vancouver, British Columbia V6T 1Z1, Canada}
\author{Jonathan Lavoie}
\affiliation{Institute for Quantum Computing and Department of Physics and
Astronomy, University of Waterloo, Waterloo, Ontario N2L 3G1, Canada}
\author{Rainer Kaltenbaek}
\affiliation{Institute for Quantum Computing and Department of
Physics and Astronomy, University of Waterloo, Waterloo, Ontario N2L
3G1, Canada} \affiliation{Faculty of Physics, University of Vienna,
Boltzmanngasse 5, A-1090 Vienna, Austria}
  \date{\today}

\begin{abstract}
Bound entangled states are the exotic objects in the entangled world. They
require entanglement to create them, but once they are formed, it is not
possible to locally distill any free entanglement from them. It is only until
recently that a few bound entangled states were realized in the laboratory.
Motivated by these experiments, we propose schemes for creating various
classes of bound entangled states with photon polarization. These include
Ac\'in-Bru\ss-Lewenstein-Sanpara states, D\"ur's states, Lee-Lee-Kim bound
entangled states, and an unextendible-product-basis bound entangled state.
\end{abstract}
 \maketitle

\section{Introduction}
Entanglement, arguably, one of the most important ingredients in
quantum information, has been extensively explored and investigated
in recent years~\cite{Entanglement}. It can exist in the form of
pure entangled states such as Bell states, which enable many quantum
information processing protocols~\cite{NielsenChuang00}. In the
regime of mixed states, entanglement can exhibit more
features~\cite{Entanglement}. Of these, the bound entangled
states~\cite{Horodecki} have very fragile entanglement properties;
they require entanglement to create them, but once they are formed,
there is no way of locally distilling any useful free entanglement
from them. Bound entangled states lie on the border of entangled
states with un-entangled states. While the characteristics of bound
entanglement are interesting theoretically  in their own right, they
are usually considered to be of little practical use, analogous to
heat in thermodynamics. Nevertheless, some bound entangled states
have found application in information concentration~\cite{Murao},
bi-partite activation~\cite{activation}, multi-partite
superactiviation~\cite{super} and secure key
distillation~\cite{SecureKey,Augusiak}, as well as providing a
resource for certain zero-capacity quantum channels~\cite{Smith}. It
is likely that more applications with bound entanglement will be
found. Nevertheless, it was not until very recently that the
synthesis of certain bound entangled states was attempted in the
laboratory, including multi-qubit
states~\cite{Amselem,Kevin,Barreiro,pseudo} and continuous-variable
states~\cite{continuous}.

This paper considers a collection of multi-qubit bound entangled
states of various types, including Ac\'in-Bru\ss-Lewenstein-Sanpara
three-qubit states~\cite{Acin}, Smolin's four-qubit
state~\cite{Smolin}, D\"ur's $N$-qubit states~\cite{Dur},
Lee-Lee-Kim $N$-qubit bound entangled states~\cite{LeeLeeKim} and an
unextendible-product-basis (UPB) bound entangled state~\cite{UPB}.
Motivated by recent experiments on creating bound entangled
states~\cite{Amselem,Kevin,Barreiro,pseudo,continuous}, we propose
schemes for creating these states using photon polarization.
\section{Smolin's bound entangled state}
This section gives a brief review of the ideas in
Refs.~\cite{Amselem,Kevin}; see also Ref.~\cite{HorodeckiReview}.
The key point is that mixed states are created by a statistical
mixture of pure states.

The Smolin state~\cite{Smolin} is a four-qubit mixed state
\begin{equation}
\label{eqn:rhoABCD} \rho^{ABCD}_{\rm
Smolin}\equiv\frac{1}{4}\sum_{i=0}^3\big(\ketbra{\Psi_i}\big)_{\rm AB}\otimes
\big(\ketbra{\Psi_i}\big)_{\rm CD},
\end{equation}
where the $\ket{\Psi}$'s are the four Bell states
$\ket{\Phi^\pm}=(\ket{00}\pm\ket{11})/\sqrt{2}$ and
$\ket{\Psi^\pm}=(\ket{01}\pm\ket{10})/\sqrt{2}$. The
form~(\ref{eqn:rhoABCD}) suggests that  the Smolin state can be
generated by mixing the four terms of the four-particle states, each
being a product of two Bell pairs. Note that
$\ket{\Phi^-}=\openone\otimes \sigma_z \ket{\Phi^+}$,
$\ket{\Psi^+}=\openone\otimes\sigma_x\ket{\Phi^+}$, and
$\ket{\Psi^-}=-i \openone\otimes\sigma_y\ket{\Phi^+}$ (where
$\openone$ and $\sigma$'s stand for the identity and the three Pauli
operators, respectively); namely, the four Bell states can be
locally converted into one another. Once two pairs of $\ket{\Phi^+}$
states have been produced using, e.g.,
downconversion~\cite{downconversion}, one only needs to randomly and
simultaneously apply either one of $\openone$, $\sigma_z$,
$\sigma_x$ and $\sigma_y$ with equal probability to one photon of
each entangled pair.  The resultant statistical mixture will be the
Smolin state, as was done in Refs.~\cite{Amselem,Kevin}

We remark that the entanglement in the Smolin state can be unlocked
if two of the parties can perform a joint Bell-state
analysis~\cite{Smolin}. This state can be used, e.g., in information
concentration~\cite{Murao} and multipartite
superactiviation~\cite{super}. As far as the amount of entanglement
is concerned, for the Smolin state the negativity ${\cal
N}$~\cite{Entanglement,negativity}
 is zero for any 2:2 partitioning, e.g., \{AB : CD\}, but nonzero
for 1:3 partitioning, e.g.,\{A:BCD\}. Specifically, ${\cal N}_{\rm
A:BCD}=1$ but ${\cal N}_{\rm AB:CD}=0$. Furthermore, the Smolin
state has the same amount of entanglement as the
Greenberger-Horne-Zeilinger-(GHZ) state, as quantified by the
geometric measure of entanglement~\cite{Wei04,GE}.

Note that the state $\rho^{ABCD}_{\rm Smolin}$ can  also be written as
\begin{equation}
\label{eqn:rhoABCD2} \rho^{ABCD}_{\rm
Smolin}=\frac{1}{4}\sum_{i=0}^3\ketbra{\Chi_i},
\end{equation}
where the $\ket{\Chi}$'s are the four orthogonal GHZ states:
\begin{eqnarray}
&&\!\!\!\!\!\!
\ket{\Chi_0}\!\equiv\!\frac{1}{\sqrt{2}}\big(\ket{0000}\!+\!\ket{1111}\big),
\
\ket{\Chi_1}\!\equiv\!\frac{1}{\sqrt{2}}\big(\ket{0011}\!+\!\ket{1100}\big),\nonumber\\
&&\!\!\!\!\!\!
\ket{\Chi_2}\!\equiv\!\frac{1}{\sqrt{2}}\big(\ket{0101}\!+\!\ket{1010}\big), 
\
\ket{\Chi_3}\!\equiv\!\frac{1}{\sqrt{2}}\big(\ket{0110}\!+\!\ket{1001}\big).\nonumber
\end{eqnarray}
This also provides an alternative way of creating the Smolin state
using a four-photon GHZ state as a resource. In practice, the
four-photon GHZ state is also created from two pairs of Bell
states~\cite{4photon}.
 For many other states below it is essential
to use a GHZ state as a resource state. We remark that Barreiro et
al. synthesized this bound entangled state with trapped ions, first
by creating an appropriate diagonal state, followed by a four-qubit
unitary entangling gate~\cite{Barreiro}.

\section{Ac\'in-Bru\ss-Lewenstein-Sanpera bound entangled states}
Ac\'in, Bru\ss, Lewenstein and Sanpera (ABLS) have proposed a class of
three-qubit bound entangled states~\cite{Acin}, described by
\begin{eqnarray}
\rho_{\rm ABLS}(a,b,c)&=&\frac{1}{n}\left(2\ketbra{{\rm 3GHZ}}+c
\ketbra{001}+\frac{1}{c}\ketbra{110}+b\ketbra{010}\right.\nonumber\\
&&\left.+\frac{1}{b}\ketbra{101}+a\ketbra{100}+\frac{1}{a}\ketbra{011}\right),
\end{eqnarray}
where $\ket{\rm 3GHZ}$ is the three-qubit GHZ state $\ket{\rm
3GHZ}\equiv(\ket{000}+\ket{111})/\sqrt{2}$ and the parameter $n$ is
a normalization factor $n\equiv 2+a +1/a+b+1/b+c+1/c$, and
parameters $a$, $b$ and $c$ satisfy $a,b,c>0$ and $abc\ne 1$. The
last condition, derived using the so-called range
criterion~\cite{Entanglement}, ensures that the ABLS state is
entangled~\cite{Acin}. This family of states can be rewritten as

\begin{equation}
\rho_{\rm ABLS}(a,b,c)=\frac{1}{n}\left(2\ketbra{{\rm
3GHZ}}+\rho_a^{(+)}+\rho_a^{(-)}+\rho_b^{(+)}+\rho_b^{(-)}+\rho_c^{(+)}+\rho_c^{(-)}\right),
\end{equation}
where the un-normalized states $\rho_{a,b,c}^{(\pm)}\equiv
\ketbra{\psi_{a,b,c}^{(\pm)}}$'s are defined via the following un-normalized
states,
\begin{eqnarray}
\ket{\psi_c^{(\pm)}}&\equiv &\sqrt{\frac{c}{2}}\ket{001}\pm\frac{1}{\sqrt{2c}}\ket{110} \\
\ket{\psi_b^{(\pm)}}&\equiv &\sqrt{\frac{b}{2}}\ket{010}\pm\frac{1}{\sqrt{2b}}\ket{101} \\
\ket{\psi_a^{(\pm)}}&\equiv
&\sqrt{\frac{a}{2}}\ket{100}\pm\frac{1}{\sqrt{2a}}\ket{011}.
\end{eqnarray}

In addition to using the range criteria~\cite{Acin,Entanglement},
the existence of entanglement in this family of states can also be
detected by entanglement witnesses~\cite{Hyllus}.  That the states
are positive under partial transpose (PPT) ascertains that they are
undistillable with respect to any bipartition ~\cite{Acin}. Being
both entangled and undistillable, the above family of states are
therefore bound entangled. Kampermann et al. used a liquid-NMR
system to implement these states, and they referred to the resultant
states as pseudo bound-entangled states as the true states created
are mixture of a small relative amount of these bound entangled
states with a large amount of the completely mixed state. Strictly
speaking, no true entanglement is present in such a system, unless
the temperature is very low. It is thus interesting to see whether
these bound entangled states can be created in other systems, where
genuine entanglement can be easily achieved. Although ABLS states
were only implemented in liquid NMR, other states, such as the
Smolin's state~\cite{Smolin}, have been created in other systems,
such as photons~\cite{Amselem,Kevin} and trapped
ions~\cite{Barreiro}. The goal of this section is to provide a
scheme for creating ABLS bound entangled states with photon
polarization.

\begin{figure}
\includegraphics[width=0.5\textwidth]{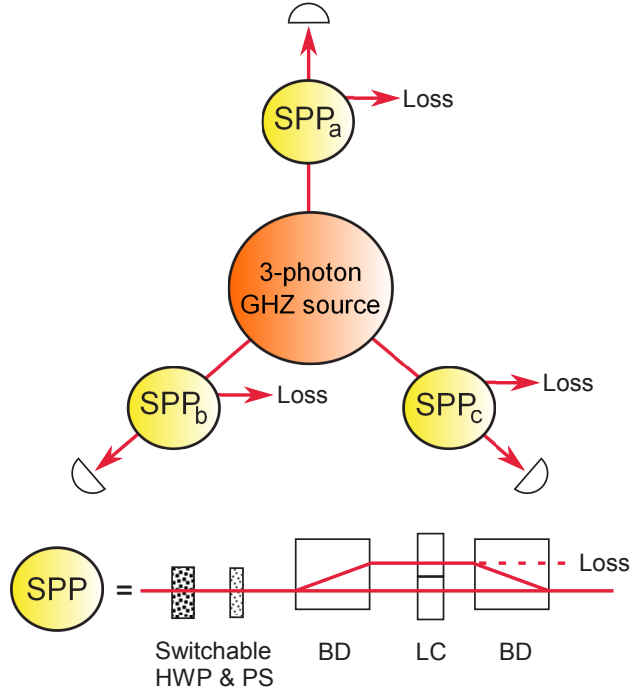}
\caption{(color online)  Scheme for creating ABLS bound entangled
states from a 3-photon GHZ state. $SPP$'s indicate collectively
either (1) identity or (2) unitary gates ($\sigma_x$ or
$\sigma_x\sigma_z$) implemented by waveplates, followed by
switchable partial polarizers. The indicated loss is due to partial
polarizers. HWP stands for the half-wave plate and PS stands for the
phase shifter that turns $V\rightarrow -V$; both elements can be
switched on and off. BD stands for the beam-displacer (see, e.g.,
Refs~\cite{Jeffrey04,Peters}), which separates, say, $H$ and $V$
polarizations, and LC stands for the liquid crystal, which acts as a
waveplate. The second BD is placed upside down so as to combine the
displaced beams. The LCs control the degree of partial polarization,
and their action can be switched on and off.} \label{fig:ABLS}
\end{figure}

Kamperman et al. synthesized their bound entangled state by first
creating an appropriate diagonal state, followed by a three-qubit
unitary gate~\cite{pseudo}. This approach is very difficult with
photons, as entangling gates are hard to come by. So how can one
create ABLS bound entangled states using photon polarization states?
Suppose we have a GHZ state $(\ket{HHH}+\ket{VVV})/\sqrt{2}$ (where
H stands for a horizontally polarized photon, and V is a vertically
polarized photon), and, for example, we have a switchable unitary
gate (implementable by Pockels cell~\cite{Jeffrey04} or liquid
crystals~\cite{Kevin}) at the path of photon 3 to control the
possible actions: (i) $H\rightarrow V$ and $V\rightarrow H$; (ii)
$H\rightarrow V$ and $V\rightarrow - H$; (iii)  do nothing, then the
resultant state is
 (i) $(\ket{HHV}+\ket{VVH})/\sqrt{2}$; (ii)
$(\ket{HHV}-\ket{VVH})/\sqrt{2}$; or (iii)
$(\ket{HHH}+\ket{VVV})/\sqrt{2}$, respectively. Furthermore, if we
have a switchable partial polarizer~\cite{Peters} (which is switched
on only in the former two cases) acting on photon 3, with
polarization-dependent transmissions being $T_H^{(c)}/T_V^{(c)}$,
then the state $(\ket{HHV}\pm\ket{VVH})/\sqrt{2}$ is transformed
(probabilistically) to
$\sqrt{T_V^{(c)}/2}\ket{HHV}\pm\sqrt{T_H^{(c)}/2}\ket{VVH}$
(un-normalized). This is the key step to create the
$\ket{\psi_c^{(\pm)}}$ state. Note that $\ket{\psi_c^{(+)}}$ and
$\ket{\psi_c^{(-)}}$ will be created with equal probability and that
one should choose $T_V^{(c)}/T_H^{(c)}=c^2$. Such partial polarizers
are an important ingredient in implementing general local filtering
operations~\cite{Peters,filtering}, as was used in various places,
such as the Procrustean method for entanglement
distillation~\cite{procrustean,Peters} and the construction of
optimal witnesses~\cite{Park}. Similarly, we can have such sets of
devices (i.e., switchable waveplates and partial polarizers)  placed
in the path of the other two photons with transmission coefficients
$T_{H/V}^{(a)}$ and $T_{H/V}^{(b)}$ (which satisfy
$T_V^{(a)}/T_H^{(a)}=a^2$ and $T_V^{(b)}/T_H^{(b)}=b^2$),
respectively, to create $\ket{\psi_a^{(\pm)}}$ and
$\ket{\psi_b^{(\pm)}}$. By appropriately mixing these states
together with the GHZ state, i.e., firing the three Pockels cells
probabilistically to match the relative weight in the state
$\rho(a,b,c)$, we will create the state at the collection output
ports, conditioned on the occurrence of a three-fold coincidence.
See Fig.~\ref{fig:ABLS} for the schematic setup. In particular, the
probabilities that no Pockels cell fires and that either $a$-, $b$-,
or $c$-th Pockels cell fires are $p_{\rm GHZ}$, $p_a$, $p_b$, or
$p_c$, respectively, and they should satisfy
\begin{equation}
p_{\rm GHZ}:p_a T_V^{(a)}:p_b T_V^{(b)}:p_c T_V^{(c)}=2:a:b:c.
\end{equation}

\section{D\"ur-Cirac states and derived bound entangled states}
It was shown by D\"ur and Cirac~\cite{DurCirac} that an arbitrary $N$-qubit
state $\rho$ can be locally depolarized into the form
\begin{eqnarray}
\rho_{\rm DC}& =&\lambda_0^+\ketbra{\Psi^+_0}+\lambda_0^-\ketbra{\Psi^-_0}\nonumber \\
&+&
\sum_{j=1}^{2^{N\!-\!1}-1}\lambda_j\big(\ketbra{\Psi^+_j}+\ketbra{\Psi^-_j}\big),\label{eqn:DC}
\end{eqnarray}
while preserving $\lambda_0^\pm=\langle\Psi_0^\pm|\rho|\Psi_0^\pm\rangle$ and
$\lambda_j=\langle\Psi_j^+|\rho|\Psi_j^+\rangle+\langle\Psi_j^-|\rho|\Psi_j^-\rangle$,
where $\ket{\Psi^\pm_0}\equiv (\ket{0^{\otimes N}}\pm\ket{1^{\otimes
N}})/\sqrt{2}$, and the $\ket{\Psi^\pm_j}$'s are GHZ-like states
\begin{equation}
\label{eqn:Belllike}
\ket{\Psi^\pm_j}\equiv(\ket{\underline{j},0}\pm\ket{\underline{2^{N-1}-j-1},1})/{\sqrt{2}}=(\ket{j_1j_2\dots
j_{N-1}0}\pm\ket{\bar{j}_1\bar{j}_2\dots\bar{j}_{N-1}1})/\sqrt{2},
\end{equation}
where $j=1,\ldots, 2^{N-1}-1$, $j_1j_2\dots j_{N-1}$ is the binary
representation of $j$ with $j_k=0$ or $1$, and $\bar{j}_k\equiv
1-j_k$. Normalization gives the condition
\begin{equation}
\lambda_0^++\lambda_0^-+2\sum_{j\ne 0}\lambda_j=1.
\end{equation}

Now define $\Delta\equiv \lambda_0^+-\lambda_0^-$, which we assume
to be non-negative without loss of generality. Consider a bipartite
partitioning $I_j$ ($j\ne 0$) which divides $1,2,..,N$ into two
groups, with one of them containing indices $k$ such that $j_k=1$ in
the binary representation of $j=j_1j_2...j_{N-1}$. The rest of the
indices $m$ (with $j_m = 0$), in addition to $N$, are contained in
the other group. One can compute the negativity with respect to the
partitioning $I_j$, and obtains~\cite{DurCirac}
\begin{equation}
\label{eqn:NegDurCirac} {\cal N}_{I_j}=\max\{0,\Delta
-2\lambda_{j}\}.
\end{equation}
A sufficient condition to infer that the state is entangled is
${\cal N}_{I_j}>0$ for certain $j$, and this means that
$\Delta>2\lambda_j$. On the other hand, when ${\cal N}_{I_j}=0$,
i.e.,\begin{equation} \label{eqn:nodistill}
 2\lambda_j\ge\Delta,\nonumber
\end{equation}
this condition implies that there cannot exist distillable
entanglement across the bipartition $I_j$~\cite{DurCirac}.

 From the form of the states~(\ref{eqn:DC}), it is easy to see that
all the above D\"ur-Cirac states can be created by the method of
mixing, once an $N$-partite GHZ state can be generated. In the
following we shall discuss D\"ur and  Lee-Lee-Kim bound entangled
states and their generalization. All these belong to the class of
D\"ur-Cirac states.

\subsection{D\"ur's bound entangled states}
D\"ur~\cite{Dur} found that for
$N\ge 4$ the following state is bound entangled:
\begin{equation}
\rho_{\rm
D}\equiv\frac{1}{N+1}\left(\ketbra{\Psi_G}+\frac{1}{2}\sum_{k=1}^N\big(P_k+\bar{P}_k\big)\right),
\end{equation}
where $\ket{\Psi_G}\equiv \big(\ket{0^{\otimes N}}+\ket{1^{\otimes
N}}\big)/ {\sqrt{2}}$ is an $N$-partite GHZ state;
$P_k\equiv\ketbra{u_k}$ is a projector onto the state
$\ket{u_k}\equiv\ket{0}_1\ket{0}_2\ldots\ket{1}_k\ldots\ket{0}_N$;
and $\bar{P}_k\equiv\ketbra{v_k}$ projects on to
$\ket{v_k}\equiv\ket{1}_1\ket{1}_2\ldots\ket{0}_k\ldots\ket{1}_N$.
It has been shown that this state violates  the Mermin-Klyshko-Bell
inequality for $N\ge 8$~\cite{Dur}, and that it violates   a
three-setting Bell inequality for $N\ge 7$~\cite{Kwek02} and a
functional Bell inequality for $N\ge 6$~\cite{SenSenZukowski02}.
From the experimental point of view, it is better to have a range of
parameters that the bound entangled states reside in, as this
results in less stringent requirements on the experimental
errors~\cite{NatPhysComment}. Indeed, it was shown in
Ref.~\cite{Wei04} that for $N\ge 4$ the family of states
\begin{equation}
\rho_{\rm D}(x)\equiv x\ketbra{\Psi_G}+\frac{1-x}{2N}
\sum_{k=1}^N\big(P_k+\bar{P}_k\big),
\end{equation}
are bound entangled if $0<x\le 1/(N+1)$ and is still entangled but not bound
entangled for $x > 1/(N+1)$. This can be seen from the fact that the
negativities of $\rho_{\rm D}(x)$ with respect to the two different partitions
$(1:2\cdots N)$ and $(12:3\cdots N)$ are
\begin{subequations}
\begin{eqnarray}
&&\!\!\!\!\!\!\!\!\!\!{\cal N}_{1:2\cdots
 N}\big(\rho_{\rm D}(x)\big)=\max\left\{0,[{(N\!+\!1)\,x-1}\,]/{N}\,\right\}, \\
 &&\!\!\!\!\!\!\!\!\!\!{\cal N}_{12:3\cdots
 N}\big(\rho_{\rm D}(x)\big)=x.
\end{eqnarray}
\end{subequations}

In contrast to Smolin's state, the four-qubit D\"ur's states have
zero negativity with respect to a 1:234 partition but non-zero
negativity with respect to a 12:34 partition.

\begin{figure}
\includegraphics[width=0.5\textwidth]{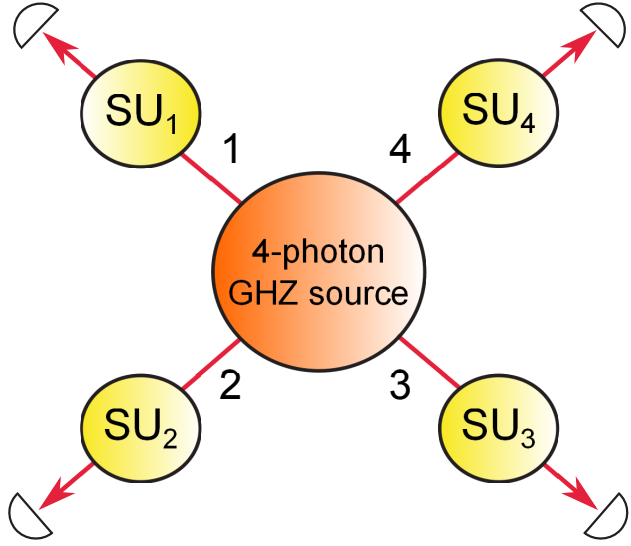}
\caption{(color online)  Scheme for creating D\"ur-Cirac states,
D\"ur bound entangled states, LLK bound entangled states. It is
illustrated with a 4-photon GHZ state source. $SU$'s represent
 unitary gates that allow to switch between
$\openone$, $\sigma_x$ and $\sigma_x\sigma_z$.} \label{fig:DurCirac}
\end{figure}

Instead of the original form by D\"ur~\cite{Dur}, we can rewrite
$P_k+\bar{P}_k$ as a mixture of GHZ-like states as follows,
\begin{equation}
P_k+\bar{P}_k=\ketbra{G_k^+}+\ketbra{G_k^-},
\end{equation}
where
\begin{equation}
\ket{G_k^\pm}\equiv
\frac{1}{\sqrt{2}}(\ket{u_k}\pm\ket{v_k})=\sigma_k^x(\sigma_k^z)^{0/1}\ket{\Psi_G},
\end{equation}
where $0$ ($1$) in the exponent of $\sigma_k^z$ corresponds to $+$
($-$). Therefore, we can also rewrite $\rho_{\rm D}(x)$ as
\begin{equation}
\rho_{\rm D}(x)= x\ketbra{\Psi_G}+\frac{1-x}{2N}
\sum_{k=1}^N\sum_{\alpha=\pm}\ketbra{G_k^\alpha},
\end{equation}
which suggests a way to create this state by mixing up various GHZ-like
states, with the probabilities being the corresponding coefficients. See
Fig.~\ref{fig:DurCirac} for the schematic setup.

 D\"ur's bound entangled states and their generalization belong
to the general D\"ur-Cirac states. In particular, in terms of
Eq.~(\ref{eqn:DC}) D\"ur's state has
$\lambda_0^+=1/(N+1),\lambda_0^-=0$, and $\lambda_j=1/2(N+1)$ for
$j=2^0, 2^1, ..., 2^{N-1}$, and zero otherwise. We remark that
photonic GHZ states of three~\cite{3photon}, four~\cite{4photon},
five~\cite{5photon} and six photons~\cite{6photon} have all been
created with high fidelity, and thus it is  within the reach of
current technology to implement  (generalized) D\"ur states, as well
as all other D\"ur-Cirac states, with up to six photons.
\subsection{Lee-Lee-Kim bound entangled states}
Lee, Lee and Kim (LLK)~\cite{LeeLeeKim} constructed bound entangled
states that are analogous to D\"ur's states (but with a different
set of $\lambda$'s)
\begin{eqnarray}
\rho_{\rm LLK}& =&\lambda_0^+\ketbra{\Psi^+_0}+\lambda_0^-\ketbra{\Psi^-_0}\nonumber \\
&+&
\sum_{j=1}^{2^{N\!-\!1}-1}\lambda_j\big(\ketbra{\Psi^+_j}+\ketbra{\Psi^-_j}\big),
\end{eqnarray}
with $\lambda_0^+=1/(N-1),\lambda_0^-=0$, and $\lambda_j=1/2(N-1)$ if $j\in
J_N\equiv\{3,6,\dots, 3\times 2^{N-3}\}$ and $\lambda_j=0$ otherwise.

It is easy to see that the LLK states are entangled as they have the
negativity~(\ref{eqn:NegDurCirac}) ${\cal N}=2/(N-1)$ with respect
to the partition $\{1,2,...,N-1: N\}$. It turns out that the
non-distillability conditions~(\ref{eqn:nodistill}) covered by the
various partitionings induced by $j\in J_N$ are sufficient to
conclude that the states are non-distillable and hence bound
entangled~\cite{LeeLeeKim}. Lee, Lee and Kim also showed that for
$N\ge 6$, the state violates the Mermin-Klyshko-Bell
inequality~\cite{LeeLeeKim}.

Similar to D\"ur's states, we generalize the parameter range of the
LLK states:
\begin{eqnarray}
\rho_{\rm LLK}(x) \equiv x\ketbra{\Psi^+_0}+ \frac{1-x}{2(N-2)}\sum_{j\in
j_N}\big(\ketbra{\Psi^+_j}+\ketbra{\Psi^-_j}\big).
\end{eqnarray}
They are bound entangled for $0<x\le 1/(N-1)$.

\begin{figure}
\includegraphics[width=0.5\textwidth]{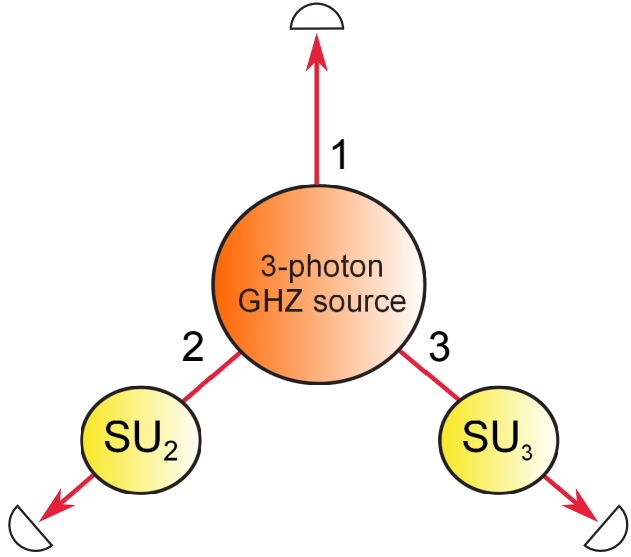}
\caption{(color online)  Scheme for creating the Chi et al.
three-qubit bound entangled state from a 3-photon GHZ state. $SU$'s
represent a unitary gates, switchable between $\openone$, $\sigma_x$
and $\sigma_x\sigma_z$.} \label{fig:ChiState}
\end{figure}
 Chi et al. later showed that for a sufficiently large number $M$ of the settings in measurement, the LLK
bound entangled states violate an $M$-setting Bell inequality if and
only if $N\ge 4$~\cite{Chi}. For $N=3$ they instead constructed a
simple three-qubit bound entangled state~\cite{Chi}
\begin{eqnarray}
\rho_3 =\frac{1}{3}\ketbra{\Psi^+_0}+
\frac{1}{6}\sum_{j=1,3}\big(\ketbra{\Psi^+_j}+\ketbra{\Psi^-_j}\big),
\end{eqnarray}
which violates not a Bell inequality but a  positive partial transpose
inequality, i.e, $|{\rm tr}\rho PT_N|\le 1$ with
${PT_N}\equiv2^{N-1}(\ketbra{\Psi_0^+}-\ketbra{\Psi_0^-})$~\cite{Nagata}. For
this state, $|{\rm tr}\rho_3 PT_N|=4/3$. This state can be generalized to
\begin{eqnarray}
\rho_3(x) ={x}\ketbra{\Psi^+_0}+
\frac{1-x}{4}\sum_{j=1,3}\big(\ketbra{\Psi^+_j}+\ketbra{\Psi^-_j}\big),
\end{eqnarray}
such that it is still bound entangled for $0<x\le 1/3$. Due to the
small number of qubits and the small number in the constituent pure
states, this particular state is very easy to create; see
Fig.~\ref{fig:ChiState} for a schematic setup.

\section{A bound entangled state from an unextendible product basis}
Our example of an unextendible product basis (UPB) involves three
qubits, A, B, and C: $\ket{\psi_1}\equiv\ket{0,0,0}$,
$\ket{\psi_2}\equiv\ket{1,+,-}$, $\ket{\psi_3}\equiv\ket{-,1,+}$,
and $\ket{\psi_4}\equiv\ket{+,-,1}$, where
$\ket{\pm}\equiv(\ket{0}\pm\ket{1})/\sqrt{2}$. A simple analysis
shows that there does not exist another linearly-independent {\it
product} state that is orthogonal to all four states $\ket{\psi_i}$.
Therefore, any state that is orthogonal to the four product basis
states must be entangled, and hence the basis is named UPB. Note
that the latter three states $\ket{\psi_2}$, $\ket{\psi_3}$
 and $\ket{\psi_4}$
are related via a periodic shift of all three parties $A\rightarrow
B\rightarrow C\rightarrow A$, and hence the basis is also called the
SHIFTS UPB. From the above properties of UPB~\cite{UPB}, Bennett et
al. showed that the following three-qubit mixed state is a bound
entangled state:
\begin{equation}
\rho_{\rm UPB}=\frac{1}{4}(\openone-\sum_{i=1}^4\ketbra{\psi_i}).
\end{equation}
This state is entangled by construction, as there cannot exist
product states in the subspace orthogonal to the UPB basis. Hence,
the state $\rho_{\rm UPB}$ cannot be expressed in terms of a mixture
of product states~\cite{UPB}. Bennett et al. also showed that the
UPB bound entagled state has the property of being
 two-way PPT and two-way separable, i.e., the entanglement across any
split into two parties is zero~\cite{UPB}. This can be understood
easily, as described in the following. First, the state is invariant
under the SHIFT operation. Second, there is one specific
decomposition of $\rho_{\rm UPB}$ into a mixture of four states
which are manifestly two-way separable (and thus PPT),
\begin{eqnarray}
\rho_{\rm UPB}&=&\frac{1}{4}\sum_{i=1}^4\ketbra{\phi_i}\\
\ket{\phi_1}&\equiv& \frac{1}{\sqrt{3}}(\ket{01}-\ket{10}+\ket{11})\ket{0}=\ket{\psi_1}\ket{0}\\
\ket{\phi_2}&\equiv&
\frac{1}{\sqrt{12}}(3\ket{00}+\ket{01}-\ket{10}+\ket{11})\ket{1}=\ket{\psi_2}\ket{1}\\
\ket{\phi_3}&\equiv& \frac{1}{\sqrt{6}}(\ket{01}+2\ket{10}+\ket{11})\ket{+}=\ket{\psi_3}\ket{+}\\
\ket{\phi_4}&\equiv&
\frac{1}{\sqrt{6}}(2\ket{01}+\ket{10}-\ket{11})\ket{-}=\ket{\psi_4}\ket{-}.
\end{eqnarray}
In the above decomposition, the state is two-way separable under AB:C, but
using the SHIFT invariant, we can conclude two-way separability under any
bi-partition. We remark that the entanglement of the above state has recently
been calculated using the geometric measure of entanglement and a generalized
concurrence~\cite{Branciard}, and it is shown that the amount of entanglement
is not small.
\begin{figure}
\includegraphics[width=0.5\textwidth]{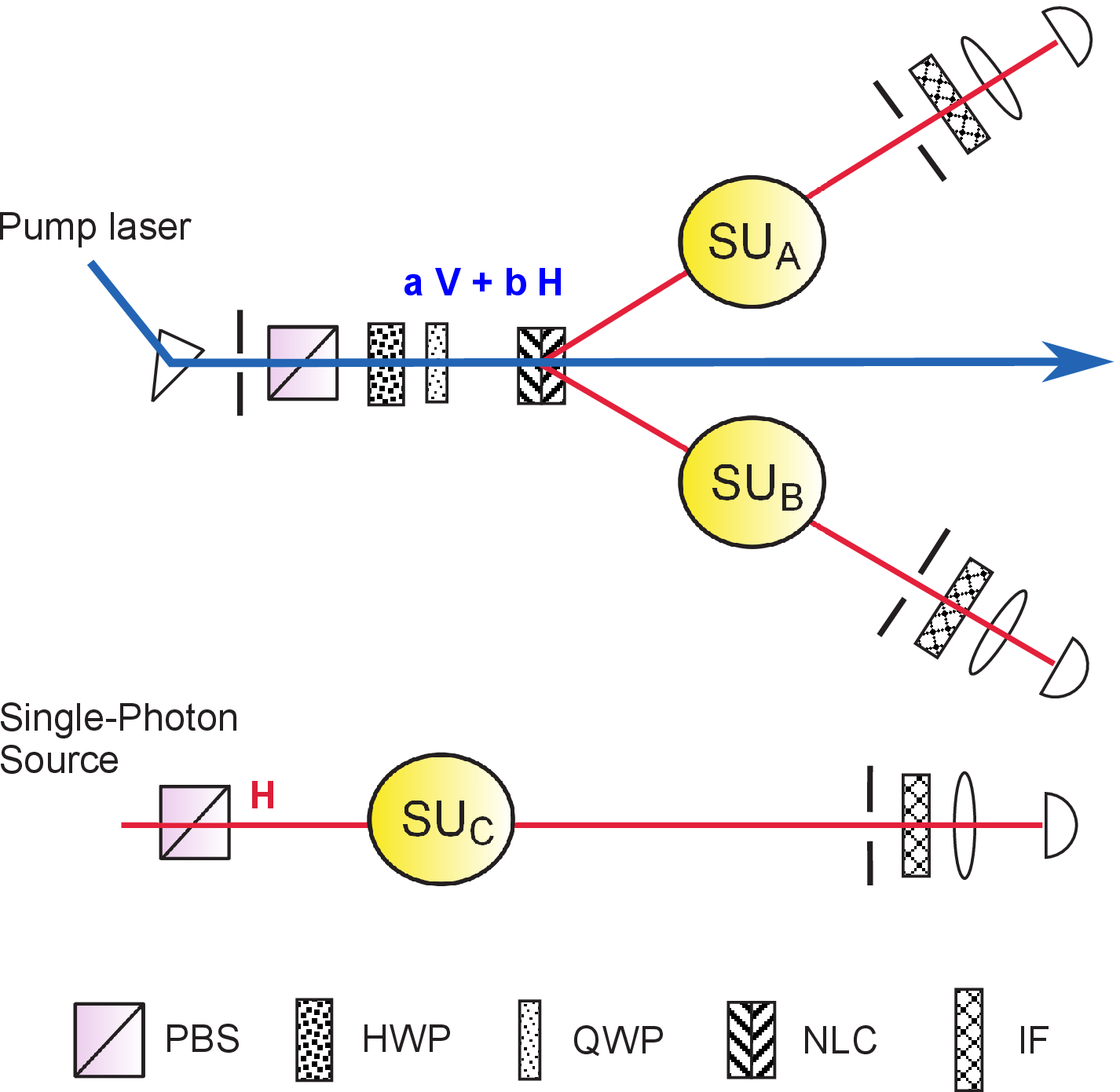}
\vspace{0.5cm} \vspace{20pt}
\includegraphics[width=0.5\textwidth]{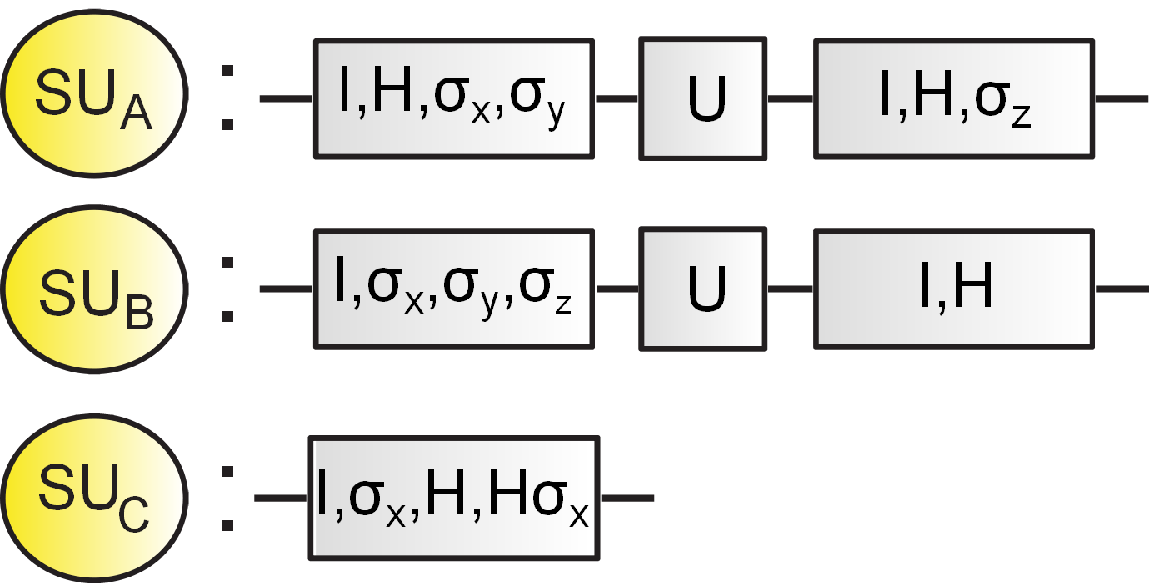}
\caption{(color online)  Scheme for creating the three-qubit UPB
bound entangled state, using a downconversion setup for generating
entangled two-photon states and a single-photon source for
generating the corresponding BB84 states. $SU$'s represent
switchable unitary gates for the respective photons. The pump
polarization state before the downconversion crystals is
$a\ket{V}+b\ket{H}$, which is then downconverted to, e.g.,
$a\ket{HH}+b\ket{VV}$ (as in the type-I process of generating
entangled pairs~\cite{downconversion}). PBS represents a polarizing
beam splitter, HWP represents a half-wave plate, QWP represents a
quarter-wave plate, NLC represents the nonlinear crystals used for
type-I downconversion, and IF represents an interference filter. For
the purpose of creating the UPB bound entangled state, the choice of
$a$ and $b$ is $a\approx 0.934$ and $b\approx 0.357$; see
Eq.~(\ref{eqn:Phi}). The exact form of the unitary gate $U$ is shown
in Eq.~(\ref{eqn:U}). The single-photon source can also be realized
by producing unentangled photon pairs via downconversion and
heralded by triggering one of the photons. } \label{fig:UPB}
\end{figure}

Since arbitrary single photon polarization states can be
created~\cite{Single} and arbitrary two-photon pure states can be
created via spontaneous parametric downconversion~\cite{Wei05}, the
above UPB bound entangled state can be created by mixing the four
constituent two-way separable three-photon pure states. The third
photon polarization appears in the four BB84 states and is
implemented by simple rotations from a fixed polarization state such
as $|H\rangle$; see Fig.~\ref{fig:UPB}.
 To create
the corresponding four two-qubit states $\ket{\psi_i}$, we first
generate a common entangled resource state $\ket{\Phi}$ from the
downconversion source, and then operate on the signal and idler
photons  by local unitaries (see the Schmidt decomposition in
Ref.~\cite{NielsenChuang00} or the Appendix in Ref.~\cite{Wei05}),
\begin{eqnarray}\
&&\ket{\psi_1}=-(U\sigma_z) \otimes (\sigma_z U) \ket{\Phi}\\
&&\ket{\psi_2}=(H U)\otimes ( U H)\ket{\Phi}\\
&&\ket{\psi_3}=(\sigma_x U H)\otimes (\sigma_x U)\ket{\Phi}\\
&&\ket{\psi_4}=(\sigma_y U)\otimes(\sigma_y U H)\ket{\Phi},
\end{eqnarray} where $H$ is the Hardamard gate, $\sigma$'s are Pauli
matrices,  $\ket{\Phi}$ is the entangled resource state and $U$ is a
single-qubit unitary gate
\begin{eqnarray}
\label{eqn:Phi}\ket{\Phi}&=&\sqrt{(3+\sqrt{5})/6}\,\ket{00}+\sqrt{(3-\sqrt{5})/6}\,\ket{11}\approx0.934172 \ket{00}+0.356822\ket{11}\\
\label{eqn:U}U&=&\begin{pmatrix}
\frac{\sqrt{5}-1}{\sqrt{10-2\sqrt{5}}}&\sqrt{\frac{2}{5-\sqrt{5}}}\cr
\sqrt{\frac{2}{5-\sqrt{5}}}&\frac{-\sqrt{5}+1}{\sqrt{10-2\sqrt{5}}}
\end{pmatrix}\approx\begin{pmatrix} 0.525731 & 0.850651 \cr 0.850651 &
-0.525731\end{pmatrix}.
\end{eqnarray}
The exact forms of $U$ and $\ket{\Phi}$ are not very illuminating and for the
actual implementation the approximate forms are sufficient. All the above
local unitaries can be implemented by waveplates~\cite{Single}. We only need
to randomly generate any of the above four states and the associated single
photon states, and the statistical mixture of the outcome will be the desired
bound entangled state. See Fig.~\ref{fig:UPB} for the schematic setup. We
remark that for the ease of implementation we have let the local unitary $U$
be always on, but other simple gates such as $\sigma_z$, $\sigma_y$,
$\sigma_x$ and $H$ need to be switched on and off depending on which state is
generated, and this can be done by Pockels cells or by liquid
crystals~\cite{Kevin}. Furthermore, the overall phase factors can be ignored,
as we are concerned with the mixture of the states.


\section{Concluding remarks}
Different physical systems, such as liquid NMR, trapped ions,
superconducting qubits, or photons may have their own preferred ways
of implementing bound entangled states, but the techniques may also
be borrowed from one another. After reviewing the schemes for
implementing Smolin's state, we have proposed schemes for creating
various classes of bound entangled states with photon polarization,
including Ac\'in-Bru\ss-Lewenstein-Sanpara states, D\"ur's states,
Lee-Lee-Kim bound entangled states, and an
unextendible-product-basis bound entangled state. These states, once
existing only in theory, can now be practically realized and tested,
e.g., via tomography or Bell inequalities in the laboratory. Some of
them turn out to be useful in information
concentration~\cite{Murao},  bi-partite
activation~\cite{activation}, multi-partite
superactiviation~\cite{super} and secure key
distillation~\cite{SecureKey}, as well as for providing a resource
for certain zero-capacity quantum channels~\cite{Smith}.

So far we have not discussed the issue of noise. Let us, for
example, consider a quantum state that undergoes the following noisy
quantum channel: $\rho\rightarrow\rho(\epsilon)= (1-\epsilon)\rho +
\epsilon \openone/ 2^N$, namely, under a depolarizing channel. The
channel only decreases entanglement content of states, and
undistillable states remain undistillable. As long as it remains
entangled, the state will still be bound entangled. Indeed, all the
bound entangled states discussed above have finite (nonzero) amount
of entanglement, as, e.g., quantified by the geometric
measure~\cite{Wei04,Branciard} or negativity across certain
partition, or by the construction of entanglement
witnesses~\cite{Hyllus}. This means that there exists a finite range
of $\epsilon$ such that $\rho(\epsilon)$ is still
entangled~\cite{Entanglement,Bill}, and hence bound entangled. This
is  a good feature for experimental implementations in order to
allow for the statistically significant observation of bound
entanglement~\cite{Kevin,NatPhysComment}.   Furthermore, even if
there are small errors in the apparatus settings,  it will in
principle be possible to apply tailored noise to overcome
experimental imperfections such that the resultant state is still
bound entangled~\cite{Kevin,Barreiro,continuous}. However, if too
much noise is introduced, the entanglement will be washed out. The
noise form that we discuss here is perhaps the simplest one. Other
forms of the noise channel can in principle be treated as well.

 We end by
remarking that the first bound entangled states were found by Horodecki in the
bi-partite systems of Hilbert spaces ${\cal C}^3\otimes {\cal C}^3$ and ${\cal
C}^2\otimes {\cal C}^4$~\cite{PHorodecki97}. We do not consider these in the
present manuscript, because they involve non-qubit systems. However, one may
consider using, e.g., two qubits to encode a three- or four-level systems, or
using other degrees of freedom, such as orbital angular momentum, as
considered in the hyperentanglement~\cite{hyper}. This is left as a possible
future direction.

\noindent {\bf Acknowledgment} T.-C.W. acknowledges useful
discussions with Julio Barreiro, Paul Kwiat, Robert Raussendorf and
Kevin Resch, as well as the hospitality of IQC, where this work was
initiated during a visit. This work was supported by NSERC and
MITACS.

\end{document}